\pdfoutput=1
\documentclass[aps,prd,amsmath,floats,floatfix, twocolumn,
superscriptaddress,nofootinbib,showpacs,longbibliography]{revtex4-1}
\usepackage{lmodern}
\usepackage[T1]{fontenc}
\usepackage[utf8]{inputenc}
\usepackage{verbatim}
\usepackage{float}
\usepackage[dvipsnames]{xcolor}
\definecolor{linkcolor}{rgb}{0.0,0.3,0.5}
\usepackage[hypertexnames=false, unicode, colorlinks=true, linkcolor=linkcolor,
citecolor=linkcolor, filecolor=linkcolor,urlcolor=linkcolor,
pdfusetitle]{hyperref}
\usepackage{hyperref}
\usepackage[all]{hypcap}
\usepackage{graphicx}
\usepackage{xspace}
\usepackage{amssymb}
\usepackage[normalem]{ulem}
\usepackage{bm}
\usepackage{orcidlink}
\usepackage{microtype}
\usepackage[english]{babel}
\usepackage{blindtext}
\usepackage{xspace}

\graphicspath{{figs/}}

\DeclareMathAlphabet{\mathpzc}{OT1}{pzc}{m}{it}

\newcommand{\RemModel}{\texttt{NRSurE\_q4NoSpin\_Remnant}\xspace}
\newcommand{\DynModel}{\texttt{NRSurE\_q4NoSpin\_Dynamics}\xspace}

\begin{document}

\title{Merger remnant and eccentricity dynamics surrogates \\for eccentric nonspinning black hole binaries}

\newcommand{\Urithiru}{\affiliation{Urithiru Tower, Mountains of Ur
            Central Roshar, Greater Hexi / Emul Border}}
\newcommand{\UMassD}{\affiliation{Department of Mathematics, Center for Scientific Computing and Data Science Research, University of Massachusetts, Dartmouth, MA 02747, USA}}
\newcommand{\AEI}{\affiliation{Max Planck Institute for Gravitational Physics(Albert Einstein Institute), D-14476 Potsdam, Germany}}
\newcommand{\KITP}{\affiliation{Kavli Institute for Theoretical Physics (KITP), University of California Santa Barbara, CA 93106, USA}}
\newcommand{\Cornell}{\affiliation{Cornell Center for Astrophysics and Planetary Science, Cornell University, Ithaca, New York 14853, USA}}
\newcommand{\CaltechB}{\affiliation{Theoretical Astrophysics, Walter Burke Institute for Theoretical Physics, California Institute of Technology, Pasadena, California 91125, USA}}
\newcommand{\Caltech}{\affiliation{TAPIR 350-17, California Institute of Technology, 1200 E California Boulevard, Pasadena, CA 91125, USA}}
\newcommand{\Olemiss}{\affiliation{Department of Physics and Astronomy, The University of Mississippi, University, MS 38677, USA}}

\newcommand{\CENTRA}{\affiliation{CENTRA, Departamento de F\'{\i}sica, Instituto Superior T\'ecnico, Universidade de Lisboa, Avenida Rovisco Pais 1, 1049-001 Lisboa, Portugal}}
\newcommand{\Radboud}{\affiliation{Department of Astrophysics/IMAPP, Radboud University Nijmegen, P.O. Box 9010, 6500 GL Nijmegen, The Netherlands}}
\newcommand{\TorontoPhysics}{\affiliation{Department of Physics 60 St.~George Street, University of Toronto, Toronto, ON M5S 3H8, Canada}} %
\newcommand{\Tokyo}{\affiliation{Research Center for the Early Universe, University of Tokyo, Tokyo, 113-0033, Japan}}%
\newcommand{\GWPAC}{\affiliation{Nicholas and Lee Begovich Center for Gravitational-Wave Physics and Astronomy, California State University Fullerton, Fullerton, California 92834, USA}} %
\newcommand{\ChristopherNewport}{\affiliation{Christopher Newport University, Newport News, VA 23606, USA}}
\newcommand{\UTAustin}{\affiliation{Department of Physics and Weinberg Institute for Theoretical Physics, University of Texas at Austin, TX 78712, USA}}
\newcommand{\ICTS}{\affiliation{International Centre for Theoretical Sciences, Tata Institute of Fundamental Research, Bangalore 560089, India}}
\newcommand{\Coimbra}{\affiliation{CFisUC, Department of Physics, University of Coimbra, 3004-516 Coimbra, Portugal}}
\newcommand{\Balears}{\affiliation{Departament de Física, Universitat de les Illes Balears, IAC3 -- IEEC, Crta.~Valldemossa km 7.5, E-07122 Palma, Spain}}
\newcommand{\KingsLondon}{\affiliation{Theoretical Particle Physics and Cosmology Group, Physics Department, King's College London, Strand, London WC2R 2LS, United Kingdom}}
\newcommand{\VSM}{\affiliation{Department of Physics, Vivekananda Satavarshiki Mahavidyalaya (affiliated to Vidyasagar University), Manikpara 721513, West Bengal, India}}
\newcommand{\Oberlin}{\affiliation{Department of Physics and Astronomy, Oberlin College}}
\newcommand{\Wigner}{\affiliation{HUN-REN Wigner RCP, H-1121 Budapest, Konkoly Thege Mikl\'{o}s \'{u}t  29-33, Hungary}}
\newcommand{\Illinois}{\affiliation{The Grainger College of Engineering, Department of Physics \& Illinois Center for Advanced Studies of the Universe, University of Illinois Urbana-Champaign, Urbana, Illinois 61801, USA}}
\newcommand{\Texas}{\affiliation{Department of Physics and Weinberg Institute for Theoretical Physics, University of Texas at Austin, TX 78712, USA}}
\newcommand{\UIB}{\affiliation{Departament de F\'isica, Universitat de les Illes Balears, IAC3 -- IEEC, Crta. Valldemossa km 7.5, E-07122 Palma, Spain}}
\newcommand{\Northwestern}{\affiliation{Department of Physics and Astronomy, Northwestern University, Evanston, IL 60208, USA}}
\newcommand{\CornellPhysics}{\affiliation{Department of Physics, Cornell University, Ithaca, NY, 14853, USA}}
\newcommand{\CornellLepp}{\affiliation{Laboratory for Elementary Particle Physics, Cornell University, Ithaca, New York 14853, USA}}
\newcommand{\CaltechSpECTRE}{\affiliation{Theoretical Astrophysics 350-17, California Institute of Technology, Pasadena, CA 91125, USA}}

\author{Adhrit Ravichandran\,\orcidlink{0000-0002-9589-3168}}
\email{aravichandran@umassd.edu}
\UMassD

\author{Peter James Nee\,\orcidlink{0000-0002-2362-5420}}
\AEI

\author{Keefe Mitman\,\orcidlink{0000-0003-0276-3856}}
\Cornell

\author{Tousif Islam}
\KITP

\author{Scott E. Field\,\orcidlink{0000-0002-6037-3277}}
\UMassD

\author{Vijay Varma\,\orcidlink{0000-0002-9994-1761}}
\UMassD

\author{Michael Boyle \orcidlink{0000-0002-5075-5116}}
\Cornell

\author{Andrea~Ceja~\orcidlink{0000-0002-1681-7299}}
\GWPAC
\Northwestern

\author{Nils Deppe \orcidlink{0000-0003-4557-4115}}
\CornellLepp
\CornellPhysics
\Cornell

\author{Noora Ghadiri\,\orcidlink{0000-0001-9162-4449}}
\Illinois

\author{Lawrence E. Kidder\,\orcidlink{0000-0001-5392-7342}}
\Cornell

\author{Prayush Kumar\,\orcidlink{0000-0001-5523-4603}}
\ICTS

\author{Marlo Morales \orcidlink{0000-0002-0593-4318}}
\GWPAC

\author{Jordan Moxon \orcidlink{0000-0001-9891-8677}}
\CaltechSpECTRE

\author{Kyle C.~Nelli \orcidlink{0000-0003-2426-8768}}
\CaltechSpECTRE

\author{Harald P. Pfeiffer\,\orcidlink{0000-0002-9994-1761}}
\AEI

\author{Antoni Ramos-Buades\,\orcidlink{0000-0002-6874-7421}}
\UIB

\author{Katie Rink \orcidlink{0000-0002-1494-3494}}
\UTAustin

\author{Hannes~R.~R\"uter\,\orcidlink{0000-0002-3442-5360}}
\CENTRA

\author{Mark A Scheel \orcidlink{0000-0001-6656-9134}}
\CaltechB

\author{Md~Arif~Shaikh\,\orcidlink{0000-0003-0826-6164}}
\VSM

\author{Leo C.\ Stein\,\orcidlink{0000-0001-7559-9597}}
\Olemiss

\author{Daniel~Tellez\orcidlink{0009-0008-7784-2528}}
\GWPAC

\author{William Throwe\,\orcidlink{0000-0001-5059-4378}}
\Cornell

\author{Nils L.\ Vu\,\orcidlink{0000-0002-5767-3949}}
\CaltechSpECTRE

\pagenumbering{arabic}
\hypersetup{pdfauthor={Ravichandran et al.}}

\date{\today}

\begin{abstract}
Accurate models of merger remnants are increasingly important for
gravitational-wave science, including precision tests of gravity with ringdown, inference of black-hole populations, and modeling hierarchical mergers. For eccentric binaries, remnant mass, spin, and recoil carry nontrivial imprints of eccentricity that are both physically informative and more challenging to model, yet remain less developed than in the quasi-circular case. We present two new models trained on numerical-relativity (NR) simulations of unequal-mass, non-spinning eccentric binary black holes: \RemModel, which predicts remnant properties, and \DynModel, a time-domain surrogate for the evolution of eccentricity and mean anomaly. Both models are trained on NR simulations over a three-dimensional parameter space with mass ratios $q \leq 4$, eccentricity $e < 0.23$, and mean anomaly $\ell \in [0,2\pi)$ radians, where both $e$ and $\ell$ defined at $t=-1000M$ relative to peak amplitude and $M$ is the total mass. We highlight some applications, including the phenomenological impact of eccentricity on remnant properties and the enhancement or suppression of recoil. We also provide error estimates for all modeled quantities, supporting reliable use in current and future gravitational-wave parameter-estimation analyses. Both models will be made available through open-source codes.
\end{abstract}

\maketitle

\section{Introduction}
\label{sec:introduction}

Since the first direct detection of gravitational waves (GWs) from a binary
black hole (BBH) merger in 2015~\cite{Abbott:2016blz}, the number of BBH
observations by the LIGO-Virgo-KAGRA (LVK) collaboration~\cite{TheLIGOScientific:2014jea, TheVirgo:2014hva, KAGRA:2020tym, Aasi:2013wya} has steadily increased, reaching $\sim200$ detections over four observing runs~\cite{LIGOScientific:2025slb}.  These observations have provided unprecedented insights into the properties of black holes and the dynamics of their mergers, which have been crucial for testing general relativity (GR) in the strong-field regime~\cite{LIGOScientific:2021sio}  and understanding the astrophysical processes that lead to the formation of BBHs~\cite{LIGOScientific:2025pvj}. In most scenarios, gravitational radiation efficiently circularizes the orbit, so  BBHs were long expected to be nearly circular by the time they enter the sensitive band of ground-based detectors. However, some astrophysical formation channels -- such as dynamical interactions in dense stellar environments~\cite{Antonini:2017ash, Fragione:2020nib, Silsbee:2016djf, Arca-Sedda:2018qgq, Vigna-Gomez:2020fvw} -- can produce BBHs that retain appreciable eccentricity in the LVK band. Detecting and characterizing such mergers would provide important constraints on BBH formation channels, their environments, and their dynamical evolution. Indeed, some candidate BBH events with possible non-negligible eccentricity have recently drawn significant attention, with multiple groups reporting tentative evidence for eccentricity in LVK observations~\cite{Ramos-Buades:2023yhy, Morras:2025xfu, Gupte:2024jfe, Clarke:2022fma, Planas:2025jny, Kacanja:2025kpr, Jan:2025fps}.

Accurate eccentric models are crucial for inference on GW datasets. If the true signal is eccentric, analyzing it with quasi-circular templates can introduce systematic parameter biases~\cite{Ramos-Buades:2019uvh}, mimic apparent deviations in tests of general relativity, and lead to incorrect conclusions about BBH populations and astrophysical formation channels. This makes it increasingly important to include eccentricity in gravitational-wave modeling, particularly as detector sensitivities improve. Because there are no closed-form solutions to the Einstein field equations for BBH systems, numerical relativity (NR) is needed to access the strong-gravity regime of eccentric comparable-mass binaries and to model the late inspiral, merger, and ringdown. However, NR is computationally expensive, making it infeasible for direct use in standard gravitational-wave data analysis~\cite{Wofford:2022ykb}.

NR surrogates are data-driven models that interpolate the results of NR simulations across the parameter space of BBH mergers. They provide a fast and accurate way to predict waveforms and remnant properties without relying on approximations based on post-Newtonian (PN) theory or black hole perturbation theory. Surrogate models for both waveforms~\cite{Blackman:2017pcm, Varma:2019csw, Islam:2021mha, Yoo:2022erv, Yoo:2023spi, Nee:2025nmh} and post-merger remnants~\cite{Varma:2019csw, Islam:2021mha, Gerosa:2018qay, Varma:2018aht, Reali:2020vkf, Islam:2023mob, Haegel:2019uop} are now widely available and routinely used in GW astronomy, with remnant surrogates also playing an important role in simulations of black hole populations in a variety of astrophysical environments~\cite{barber2025formation}.

For eccentric binaries, the NR surrogate model \texttt{NRSur2dq1EccRemnant} was previously developed for equal-mass, non-spinning eccentric merger remnants~\cite{Islam:2021mha}. In this paper, we extend that framework to unequal masses, broaden the eccentricity range, and include modeling of the recoil kick. We also introduce a companion model for the inspiral evolution of the eccentricity and mean anomaly. Because these quantities are time dependent, we adopt the following convention throughout the paper. We define the reference time $t=0M$ as the instant when the quadrature sum of all available waveform harmonic modes reaches its maximum~\footnote{As discussed later, we use both CCE and extrapolated waveform data, and their peaks can differ slightly because the mode content is not identical. CCE waveforms include modes up to $\ell \le 8$, whereas we use only the $\ell \leq 5$ modes of the extrapolated waveforms, and even the modes common to both can differ slightly. These differences are negligible for our time-alignment procedure.} (see Eq.~(24) of Ref.~\cite{Blackman:2017dfb}). All other reference times are measured relative to this choice. When we write the eccentricity or mean anomaly with a subscript, $X_{t'}$, we mean that $X$ is evaluated at time $t=t'$ under this convention. For example, $e_{-1000M}$ denotes the eccentricity at $t=-1000M$. With this notation, we introduce the following models:

\begin{itemize}
      \item \RemModel\label{name:RemnantModel}, which models the remnant mass $m_f$, spin $\chi_f$, and kick velocity vector $v_f$ of the remnant black hole formed from the merger of two unequal-mass, non-spinning, eccentric black holes. This model extends \texttt{NRSur2dq1EccRemnant}~\cite{Islam:2021mha} to mass ratios $q \leq 4$, enlarges the eccentricity range to $e_{-1000M}<0.23$, and retains dependence on the cyclic mean-anomaly parameter $\ell_{-1000M} \in [0, 2\pi)$ radians. The choice of reference times for the eccentricity and mean anomaly is explained in Sec.~\ref{subsubsection:param_ref_time}.
      \item \DynModel, which models the inspiral evolution of the eccentricity $e(t)$ and mean anomaly $\ell(t)$. It is valid for unequal masses up to $q \leq 4$, eccentricities $e_{-3000M}<0.33$, and mean anomalies $\ell_{-1200M} \in [0, 2\pi)$ radians. The larger eccentricity range reflects the fact that this model is parameterized at earlier reference times than \RemModel, and eccentricity generally decreases during the inspiral. The reference time for \DynModel was chosen to match the parameterization of the waveform surrogate \texttt{NRSurE\_q4NoSpin\_22}~\cite{Nee:2025nmh}. Although the two models use different parameterizations, they are intended to describe the same class of non-spinning eccentric binaries over the shared physical domain covered by the available NR simulations; the differences in the training sets are explained in Sec.~\ref{subsubsection:auxiliary_paramspace}.
\end{itemize}

This paper is organized as follows. In Sec. \ref{sec:methods}, we describe the methods used to build our models. In Sec. \ref{sec:results}, we assess the accuracy of the models by comparing their errors to the resolution errors of the NR simulations. In Sec. \ref{sec:applications}, we present example applications of the new models by exploring the phenomenological impact of eccentricity on various remnant properties. Finally, in Sec. \ref{sec:conclusions}, we discuss the validity of the models and outline directions for future work.

\begin{figure}[tb!] %
      \includegraphics{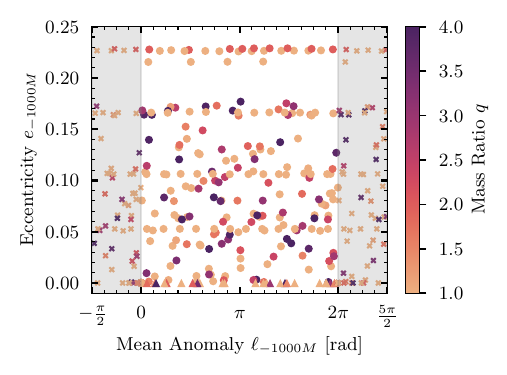}
      \caption{\RemModel three-dimensional parameter space coverage. Each circular marker represents an NR simulation. As explained in Sec.~\ref{subsubsection:periodicity_degeneracy}, crosses mark duplicated points used to enforce periodicity in $\ell$, while triangles mark duplicated points used to enforce mean-anomaly degeneracy for quasi-circular systems. The gray region denotes the extended mean-anomaly domain and is filled with crosses. The vertical (horizontal) axis shows the eccentricity (mean anomaly) at $t=-1000M$, and marker color indicates the mass ratio.}
      \label{fig:RemnantParameterSpaceDuplicated}
\end{figure}

\section{Methods}
\label{sec:methods}

Because this paper introduces two surrogate models, we present their methodologies separately. Section~\ref{subsection:remnant_methods} introduces \RemModel, including its parameter space and data-extraction procedure (Secs.~\ref{subsubsection:remnant_paramspace} and \ref{subsubsection:remnant_extrapolation_CCE}), surrogate construction (Sec.~\ref{subsubsection:remnant_surrogation}), and treatment of periodicity and mean-anomaly degeneracies (Sec.~\ref{subsubsection:periodicity_degeneracy}). Section~\ref{subsection:auxiliary_methods} then introduces \DynModel, including its parameter space and data extraction (Sec.~\ref{subsubsection:auxiliary_paramspace}) and surrogate modeling procedure (Sec.~\ref{subsubsection:auxiliary_surrogation}).
\begin{figure}[tb!]
      \includegraphics{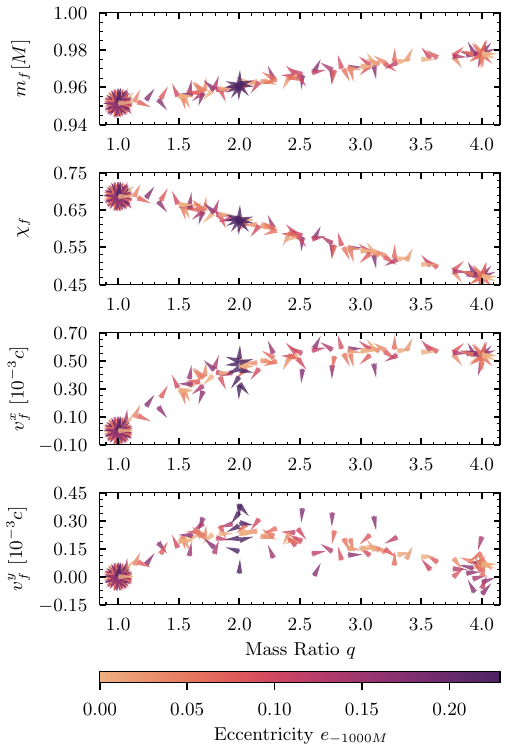}
      \caption{\RemModel properties that are modeled. Each arrow marker represents the value of (top to bottom): remnant mass, remnant spin, remnant kick velocity's $x$-component, and remnant kick velocity's $y$-component. The horizontal axis is the mass ratio, the color of the arrows represents the value of $e_{-1000M}$, and the angle of the arrow with respect to the $x$ axis represents the mean anomaly $\ell_{-1000M}$. For example, a right-pointing arrow indicates $\ell_{-1000M}=0$ radians.}
      \label{fig:RemnantProperties}
\end{figure}

\subsection{\RemModel}
\label{subsection:remnant_methods}

\subsubsection{Parameter space and properties modeled}
\label{subsubsection:remnant_paramspace}
Relaxing the quasi-circular restriction enlarges the binary black hole parameter space by two additional dimensions. One can parameterize these extra dimensions through a reference eccentricity $e$ and mean anomaly $\ell$~\cite{Shaikh:2023ypz}. The former quantifies the deviation of the orbit from a perfect circle, while the latter describes the position of the binary away from the pericenter (point of closest approach) along its eccentric orbit at a given time. There are multiple ways to define these parameters in GR~\cite{Islam:2025oiv, Vijaykumar:2024piy}, but we use a waveform-based definition following the prescription in Refs.~\cite{Shaikh:2025tae,Shaikh:2023ypz}. The eccentricity is computed by constructing interpolants through the instantaneous GW frequency through periastrons and apastron times, and using Eqs.~(4), (8) and (9) of Ref.~\cite{Shaikh:2023ypz}. The mean anomaly is a piecewise linear function, varying from $0$ to $2\pi$ between consecutive periastron times. We use the \texttt{gw\_eccentricty}\footnote{git hash c916d7032b97c13b1219c401301dc4d141e5354f} package using the \texttt{AmplitudeFits} method~\cite{Shaikh:2023ypz} for the eccentricity and mean anomaly computation.

The parameter space for this model is three-dimensional, spanning mass ratio $q \le 4$, eccentricity $e_{-1000M} < 0.23$, and mean anomaly $\ell_{-1000M} \in [0, 2\pi)$ radians, using the conventions introduced in Sec.~\ref{sec:introduction}. Figure~\ref{fig:RemnantParameterSpaceDuplicated} shows the coverage of this space where each circular marker corresponds to a numerical-relativity simulation performed with the Spectral Einstein Code (\texttt{SpEC}). Our training set consists of 203 simulations from the SXS catalog~\cite{Scheel:2025jct} with identifiers \texttt{SXS:BBH:}[1221-1222, 1354, 2331, 2483, 2485, 2570-2579, 2580-2589, 2590-2599, 2600-2609, 2610-2616, 3617, 3731-3776, 3778-3821, 4293-4302, 4304-4305, 4308-4315, 4317-4324, 4326-4332, 4357-4377, 4379-4381]. The waveform durations vary between about $3400M$ and $13400M$.

Figure~\ref{fig:RemnantProperties} depicts the remnant properties we model, which are the mass $m_f$, dimensionless spin magnitude $\chi_f$, and kick velocity $\vec{v}_f$ of the remnant BH. Nonprecessing binaries have a constant angular momentum direction, and symmetry implies that the kick lies in the orbital plane while the final spin is orthogonal to it~\cite{Boyle:2007sz}. By convention, we choose the $z-$axis to be along the orbital angular momentum direction, and therefore only model the kick components in the orbital plane (denoted as $v_f^x$ and $v_f^y$). Here, the positive $x-$axis points from the lighter BH to the heavier BH at the reference time $t=-100M$. The structure in the remnant kick distribution arises from this choice of frame.

From Fig.~\ref{fig:RemnantProperties}, it is evident that the dominant effect comes from the mass-ratio $q$, and the effects of eccentricity and mean anomaly are subdominant. Also, the effect of mean anomaly and eccentricity are stronger for the kick velocities than the remnant mass and spin. This is because the kick depends on the instantaneous linear momentum flux at coalescence, in which $e$ and $\ell$ enter at leading order (see Eqs.~(6.1) and (6.3) of Ref.~\cite{Kastha:2021kyn}), and therefore have a stronger effect on the kick velocity. The $m_f$ and $\chi_f$ depend on the instantaneous energy and angular momentum fluxes, and $e$ and $\ell$ enter their expressions at a higher order (see Eqs.~(13) and (14) of Ref.~\cite{Khairnar:2024rzs}). 

Figures~\ref{fig:RemnantParameterSpaceDuplicated} and~\ref{fig:RemnantProperties} indicate that our dataset has several points at $q=1$. This is because we re-use the NR simulations used in previous works~\cite{Islam:2021mha,Nee:2025zdy} that required targeted NR simulations on $q=1$ slice of the non-spinning eccentric parameter space.

\subsubsection{Computing remnant quantities from simulation data}
\label{subsubsection:remnant_extrapolation_CCE}
With earlier remnant surrogate models~\cite{Varma:2018aht,Varma:2019csw}, the remnant kick was obtained by computing the linear-momentum flux~\cite{Ruiz:2007yx} from extrapolated gravitational waveform data and integrating it in time to get the net radiated momentum. Dividing the net radiated momentum by the remnant mass yields the kick velocity~\cite{Gerosa:2018qay}. In this approach, waveforms are extracted on a set of finite-radius spheres and then extrapolated to future null infinity, after which center-of-mass corrections are applied~\cite{Boyle:2009vi,Scheel:2025jct,Boyle:2015nqa}.

A drawback of the computation based on the linear-momentum flux, however, is that the kick depends on the chosen integration start time. For quasi-circular systems this dependence is typically small, because the center-of-mass velocity remains close to its average value over most of the early inspiral. For eccentric systems, however, the sensitivity can be substantially larger, and can change individual kick components at the few-percent level (up to $\mathcal{O}(5\%)$), leading to noticeably worse models.

A more robust means of extracting the remnant's properties is to compute them from the Poincaré charges at future null infinity~\cite{Iozzo:2021vnq}. These charges depend on the Weyl scalars $\Psi_{1}$ and $\Psi_{2}$ which are not typically output as a part of extrapolation for NR simulations. Because of this, in this work we instead use Cauchy-characteristic evolution (CCE) to compute both the strain waveform and the Weyl scalars at future null infinity. This was done using the \texttt{SpECTRE} code's implementation of CCE~\cite{spectrecode,Moxon:2020gha,Moxon:2021gbv}.

Using CCE datasets, we follow the procedure of Refs.~\cite{GomezLopez:2017kcw, T_Dray_1985,T_Dray_1984,Streubel1978ConservedQF} to extract remnant properties from the strain and Weyl scalars. We compute the remnant mass from the Bondi rest mass and the remnant spin from the Bondi dimensionless spin charge, using Eqs.~(11) and (15) of Ref.~\cite{Iozzo:2021vnq} evaluated at the final simulation time. For the recoil, we use the center-of-mass charge (Eq.~(7.1) of Ref.~\cite{Mitman:2024uss}). Specifically, we fit this charge with degree-one polynomials over two time windows: an inspiral window $[t_{\mathrm{peak}}-2000, t_{\mathrm{peak}}-500]$ and a ringdown window $[(t_{\mathrm{final}}-t_{\mathrm{peak}})/2, t_{\mathrm{final}}-t_{\mathrm{peak}}]$, where (as before) $t_{\mathrm{peak}}$ is the time at which the $L^2$ norm of the strain on the two-sphere is maximized, and $t_{\mathrm{final}}$ is the end of the simulation. We then define the kick as the difference between the slopes of the ringdown and inspiral fits. Kick models based on this methodology were recently presented in Ref.~\cite{DaRe:2025glj}, which also compares this approach to other recoil estimates for quasi-circular systems, including momentum-flux methods with extrapolated waveforms. Ref.~\cite{DaRe:2025glj} reports CCE and extrapolation-based kicks differing by up to $\sim 5\%$ for the cases studied there. The origin of this systematic effect remains unclear.

\subsubsection{Modelling procedure}
\label{subsubsection:remnant_surrogation}
We use Gaussian process regression (GPR) to interpolate the remnant properties over the three-dimensional parameter space. GPR is a nonparametric Bayesian regression method that is well suited to modeling complex, high-dimensional datasets. In addition to predictions, it provides uncertainty estimates, which are useful for assessing the reliability of the model. We adopt the GPR setup described in Ref.~\cite{Varma:2018aht}, implemented using scikit-learn~\cite{scikit-learn}.

In addition to the GPR $1\sigma$ error estimate, we assess the accuracy of our procedure using $K$-fold cross-validation with $K=20$. We randomly divide the training dataset into $K$ mutually exclusive sets. For each set, we construct the fits using the other $K-1$ sets and test them on the held-out set. This yields out-of-sample errors that conservatively estimate the accuracy of our fits. We compare these errors with the numerical truncation error in the NR simulations, estimated from differences in the relevant remnant quantities between the two highest available resolutions.

\subsubsection{Periodicity and degeneracy in mean anomaly}
\label{subsubsection:periodicity_degeneracy}

The mean anomaly $\ell$ is a cyclic parameter, so $\ell=0$ radians and $\ell=2\pi$ radians correspond to the same physical configuration. To ensure that the model respects this symmetry, we encourage periodicity in the mean-anomaly direction by extending the domain from $[0,2\pi)$ to $[-\delta,\,2\pi+\delta)$. Here, $\delta$ is a small positive buffer chosen so that enough duplicated points are available near the boundaries for the GPR model to learn the periodic structure in $\ell$ and make accurate predictions there. For each training point $(e_i,\ell_i,q_i)$, if $\ell_i \in [0,\delta]$ we add a duplicate point at $(e_i,\ell_i+2\pi,q_i)$, and if $\ell_i \in [2\pi-\delta,2\pi]$ we add a duplicate point at $(e_i,\ell_i-2\pi,q_i)$. In each case, the duplicated point is assigned the same remnant properties as the original one.

For the remnant mass, spin, and kick, we find that $\delta=\pi/2$ works well for our dataset. We adopt this choice with one exception: for the kick velocity in equal-mass systems, we use $\delta=\pi/4$ instead. For equal masses, one expects $v_f^x=v_f^y=0$, but numerical error leaves residual values with magnitude $|v_f^{x/y}| \sim 10^{-7}$. We find that duplicating too much data dominated by this numerical noise degrades the GPR fit. One could instead impose $v_f^x=v_f^y=0$ directly for equal-mass systems, but we found that this does not improve the fit.

At $e=0$, the mean anomaly $\ell$ is not physically meaningful, since all values of $\ell$ correspond to the same quasi-circular orbit. To encode this degeneracy during GPR training, we treat systems with $e<10^{-5}$ at the start of the simulation, as reported in the \texttt{SpEC} metadata, as effectively non-eccentric. For these data points, we set $\ell=0$ radians and add duplicates at $\ell=2\pi$ radians and at one additional value of $\ell$ sampled uniformly from $(0,2\pi)$. This produces multiple realizations of the same system at $e_{-1000M}=0$ with identical remnant properties but different values of $\ell$, allowing the GPR model to learn that the remnant is independent of mean anomaly in the quasi-circular limit.

This procedure enlarges the training set so that the GPR model can learn known structure in the data, namely the periodicity of the mean anomaly and the degeneracy at $e=0$. Figure~\ref{fig:RemnantParameterSpaceDuplicated} shows the resulting parameter-space coverage of the remnant surrogate after applying this procedure. In particular, the crosses denote duplicated points used to impose periodicity, the triangles denote duplicated points used to impose the mean-anomaly degeneracy at $e_{-1000M}\approx 0$, and the gray region shows the extended mean-anomaly domain.

As a proof of principle, Appendix~\ref{section:proof_of_principle_duplication} shows that this duplication procedure helps the model learn the periodic structure in mean anomaly and reduces errors near the boundaries of the $\ell_{-1000M}$ domain.

\subsubsection{Choice of parameter reference times}
\label{subsubsection:param_ref_time}

Because the eccentricity and mean anomaly evolve during the inspiral, one must choose reference epochs at which to define them for the parametric fits. Reference~\cite{Nee:2025zdy} showed that, for equal-mass non-spinning binaries, the behavior of several merger quantities becomes significantly simpler when these reference epochs are chosen sufficiently close to merger. The intuitive reason is that, for comparable-mass systems, the dependence of many merger quantities on the mean anomaly is approximately sinusoidal. The amplitude of this sinusoidal variation depends on the eccentricity, while its phase is set by the radial phase at plunge\footnote{In the test-mass limit, this is naturally described in terms of the radial phase at the crossing of the last stable orbit~\cite{Faggioli:2025hff}.}. If the mean anomaly is specified too early in the inspiral, differences in the radial frequency across parameter space lead to substantial accumulated dephasing, making the relation between the chosen reference phase and the radial phase at plunge more complicated. Choosing the reference epoch closer to merger reduces this accumulated dephasing and therefore simplifies the mapping. A similar argument applies to the eccentricity, although the effect is weaker because the eccentricity evolves slowly during the inspiral, whereas the radial phase evolves much more rapidly, advancing by $2\pi$ radians every radial period.

To determine the reference time used in \RemModel, we constructed a set of trial models in which we varied the reference time for the eccentricity and mean anomaly independently over the range $[-3000M,-1000M]$. When varying one reference time, we held the other fixed at $-1000M$. We found that choosing reference times closer to merger generally reduced the median errors for all remnant quantities and brought them closer to the median NR resolution errors. This trend was more pronounced for the mean-anomaly reference time, especially for the remnant kick velocity. These results motivated our choice of $t_{\mathrm{ref}}=-1000M$ for both the eccentricity and mean anomaly in the final model.

\subsection{\DynModel}
\label{subsection:auxiliary_methods}

\begin{figure}[tb!]
      \includegraphics{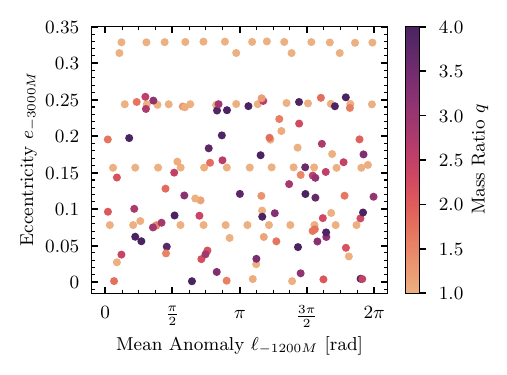}
      \caption{\DynModel parameter-space coverage. Each dot represents an NR simulation in the dataset. The coverage is less dense than in Fig.~\ref{fig:RemnantParameterSpaceDuplicated} because we restrict to simulations long enough to provide $e(t)$ and $\ell(t)$ on the time interval $[-6500M,-700M]$. This matches the time domain used by the \texttt{NRSurE\_q4NoSpin\_22} waveform model~\cite{Nee:2025nmh}.}
      \label{fig:AuxiliaryParameterSpace}
\end{figure}

\begin{figure*}[tb!]
      \includegraphics{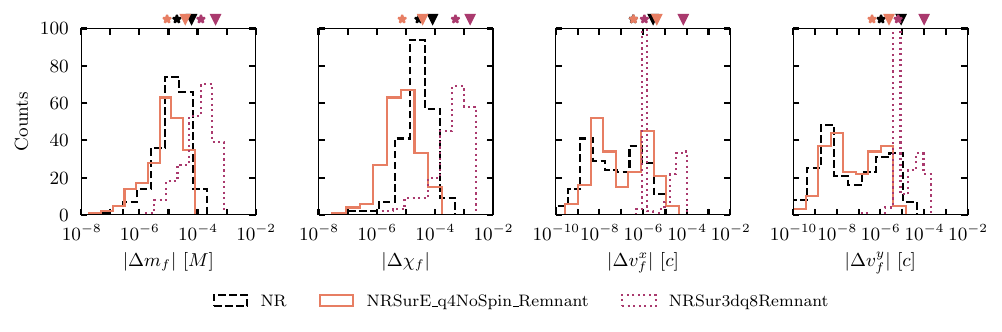}
      \caption{Error histograms for \RemModel predictions. The panels show the distributions of absolute errors in the (left to right) remnant mass, remnant spin, and remnant kick-velocity components. The orange histograms denote 20-fold cross-validation errors, and the dashed black curves denote NR resolution errors. The close agreement between them indicates that the model achieves accuracy comparable to that of the underlying NR simulations. The dotted purple lines show the errors obtained by approximating eccentric systems with the quasi-circular remnant model \texttt{NRSur3dq8Remnant} instead of \RemModel. Their larger values quantify the systematic error introduced by neglecting eccentricity and mean-anomaly effects. The star (triangle) marker depicts the median (90th percentile) error for each histogram.}
      \label{fig:RemnantSurrogateValidationHistogram}
\end{figure*}

\subsubsection{Parameter space and quantities modeled}
\label{subsubsection:auxiliary_paramspace}
The parameter space for the dynamics model is shown in Fig.~\ref{fig:AuxiliaryParameterSpace}, where each circular marker again represents an NR simulation run using the Spectral Einstein Code (\texttt{SpEC}). We use a training set of $151$ simulations from the SXS catalog~\cite{Scheel:2025jct} with identifiers: \texttt{SXS:BBH:}[3731-3747, 3749-3776, 3778-3821, 4293-4302, 4304, 4317-4324, 4326-4332, 4357-4377, 4379-4380, 4443-4455]. The dynamics model is built on a less dense parameter space compared to the remnant model, as we required our NR simulations to be long enough for the domain for $e(t)$ and $\ell(t)$ to be $[-6500M, -700M]$. The parameters are defined at reference times that match the ones used for \texttt{NRSurE\_q4NoSpin\_22}~\cite{Nee:2025nmh}. The eccentricity is defined at $t=-3000M$, while the mean anomaly is defined at $t=-1200M$. 

The quantities we model are the time-dependent eccentricity $e(t)$ and mean anomaly $\ell(t)$, extracted from the extrapolated GW strain data at extrapolation order $2$ (see Sec~2.4.1 in Ref.~\cite{Boyle:2019kee})  using the \texttt{gw\_eccentricity} package with the \texttt{AmplitudeFits} method~\cite{Shaikh:2023ypz}. Because $\ell(t)$ is cyclic, we first unwrap it to obtain a continuous time series by adding or subtracting $2\pi$ radians as needed to remove jumps between consecutive samples. We then shift each unwrapped $\ell(t)$ by an integer multiple of $2\pi$ radians so that $\ell=0$ at the periastron immediately preceding the reference time $t_{\mathrm{ref}}=-1200M$. As a result, all waveforms satisfy $\ell_{-1200M}\in[0,2\pi]$ radians. This alignment reduces variation across the dataset and reduces the number of basis functions required to represent $\ell(t)$ accurately. Since adding integer multiples of $2\pi$ radians to $\ell(t)$ does not change the binary configuration, this procedure has no effect on the waveform.

\subsubsection{Modelling procedure}
\label{subsubsection:auxiliary_surrogation}

We build the dynamics model using the surrogate algorithm described in Ref.~\cite{Field:2013cfa}. We first construct a reduced basis~\cite{Field_2011} for the training data using singular value decomposition (SVD), which yields an orthonormal basis. The basis is chosen so that the projection errors for the full dataset (see Eq.~(5) of Ref.~\cite{Blackman:2017dfb}) lie below prescribed tolerances. For our dataset, we use tolerances of $10^{-5}$ for $e(t)$ and $10^{-2}$ radians for $\ell(t)$. These values are also validated through visual inspection of the basis functions to ensure that they are not contaminated by noise. Next, we construct an empirical interpolant in time~\cite{Barrault:2004,Maday:2009,Hesthaven:2014}, which selects a set of representative time nodes. The number of empirical-interpolation nodes is equal to the number of reduced-basis functions. Finally, we use the GPR setup described in Ref.~\cite{Varma:2018aht}, implemented in scikit-learn~\cite{scikit-learn}, to interpolate the values of the time-series data at the empirical-interpolation nodes across the three-dimensional parameter space.

\begin{figure}[tb!]
      \includegraphics{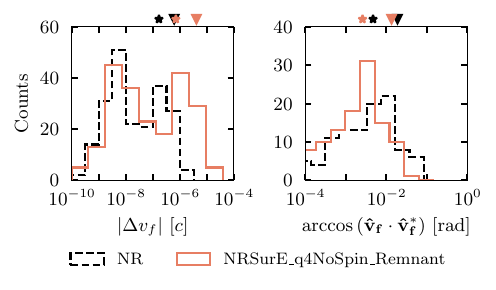}
      \caption{Error histograms for \RemModel predictions of the kick magnitude (left) and kick angle (right). The orange histograms denote $K$-fold cross-validation errors, and the dashed black curves denote NR resolution errors. The star (triangle) marker depicts the median (90th percentile) error for each histogram. In the right panel, $\mathbf{\hat{v}_{f}}$ is the kick direction from NR, while $\mathbf{\hat{v}_{f}^{*}}$ is the corresponding direction from \RemModel (orange) or from the lower-resolution NR simulation (black). The angle between these vectors defines the kick-angle error in each case. Their close agreement indicates that, although the model is trained on the kick components, it predicts the kick magnitude and angle at a level comparable to the NR simulations.}
      \label{fig:RemnantSurrogateValidationHistogramKicks}
\end{figure}

We assess the accuracy of our model using 20-fold cross validation, following the same procedure as for the remnant surrogate. The resulting out-of-sample errors provide a conservative estimate of the model accuracy. We compare these errors with the numerical resolution error in the extracted eccentricity and mean-anomaly time series, estimated by comparing $e(t)$ and $\ell(t)$ computed from the two highest-resolution NR simulations available.

\section{Results}
\label{sec:results}
\subsection{\RemModel}

Figure~\ref{fig:RemnantSurrogateValidationHistogram} shows the out-of-sample errors for \RemModel. The histograms display the distributions of absolute errors in the remnant mass $m_f$, remnant spin magnitude $\chi_f$, and remnant kick-velocity components $v_f^{x/y}$ from $20$-fold cross-validation~\footnote{As described in Sec.~\ref{subsubsection:periodicity_degeneracy}, the training data are augmented to enforce periodicity and degeneracy properties. This augmentation is applied only when building the surrogate on each training fold, so duplicated points never appear in the corresponding validation fold.}, alongside the NR resolution errors. The close agreement between the cross-validation and NR resolution errors indicates that the model is accurate to roughly the level of the NR simulations themselves. We also show the errors obtained by using the quasicircular remnant model \texttt{NRSur3dq8Remnant}~\cite{Varma:2018aht} to estimate remnant quantities for eccentric systems. Although Fig.~\ref{fig:RemnantProperties} shows that the dependence on $e$ and $\ell$ is subdominant to that on $q$, the quasicircular-model errors are between $1$ and $2$ orders of magnitude larger than both the validation and NR resolution errors, highlighting the importance of including eccentricity effects.

Additionally, Fig.~\ref{fig:RemnantSurrogateValidationHistogramKicks} shows the out-of-sample errors for the remnant kick magnitude and angle. For the kick-angle comparison, we exclude near-equal-mass systems with $q < 1.01$, since in that limit the kick magnitude approaches zero and the direction is not well defined. Although the surrogate models the kick components directly, the corresponding errors in the derived kick magnitude and angle are comparable to the NR resolution errors. This suggests that the model predicts the kick magnitude and direction with accuracy comparable to that of the underlying NR simulations.

Taken together, these results define the region of validity of the model as the portion of parameter space on which it was trained and over which it achieves accuracy comparable to that of the NR simulations themselves. For the remnant quantities computed with \RemModel, the median errors are $\approx 9 \times 10^{-6} M$ for $m_f$, $\approx 8 \times 10^{-6}$ for $\chi_f$, $\approx 4 \times 10^{-7} c$ for $v_f^x$, $\approx 4 \times 10^{-7} c$ for $v_f^y$, $\approx 7 \times 10^{-7}c $ for the kick speed, and $\approx 3 \times 10^{-3}$ radians for the kick angle.

\begin{figure*}[htp]
      \includegraphics[width=0.85\textwidth]{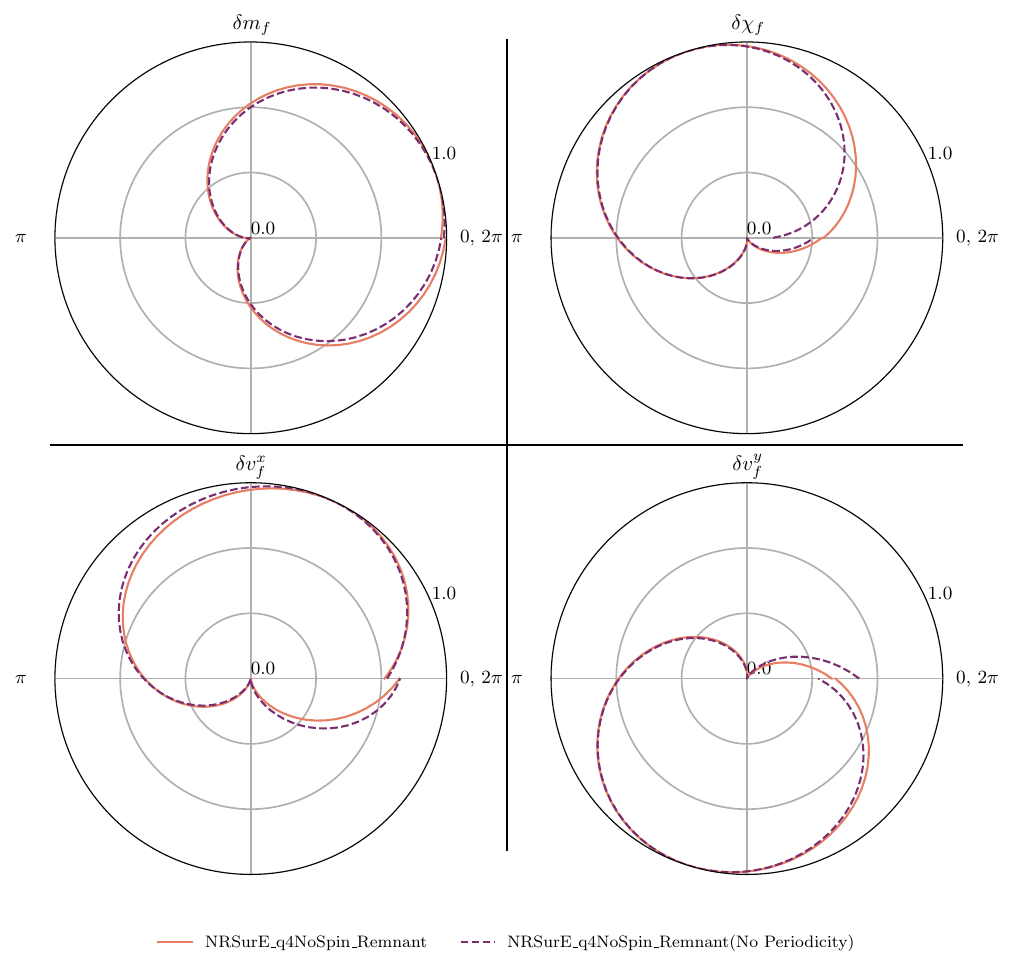}
      \caption{Evaluations of \RemModel and an alternative model, \texttt{NRSurE\_q4NoSpin\_Remnant(No Periodicity)}, which does not enforce periodicity. Both models are evaluated at fixed $q=1.5$ and $e_{-1000M}=0.15$, while $\ell_{-1000M}$ is varied over its full range. The difference is that \RemModel includes the periodicity padding described in Sec.~\ref{subsubsection:periodicity_degeneracy}, whereas \texttt{NRSurE\_q4NoSpin\_Remnant(No Periodicity)} does not. The panels show the scaled remnant mass $\delta m_f$ (top left), remnant spin $\delta \chi_f$ (top right), and remnant kick components $\delta v_f^x$ (bottom left) and $\delta v_f^y$ (bottom right). Here, $\delta X$ is defined in Eq.~\eqref{eq:rescaled_X_variation} and is plotted in polar coordinates, with polar angle $\ell_{-1000M}$ and radial coordinate given by the model output. The discontinuity between $\ell_{-1000M}=0$ and $2\pi$ is significantly reduced when periodicity padding is included, indicating that the padded model more faithfully captures the periodic dependence on $\ell_{-1000M}$.}
      \label{fig:RemnantSurrogatePeriodicityCheck_Radial}
\end{figure*}

\begin{figure*}[htb]
      \includegraphics{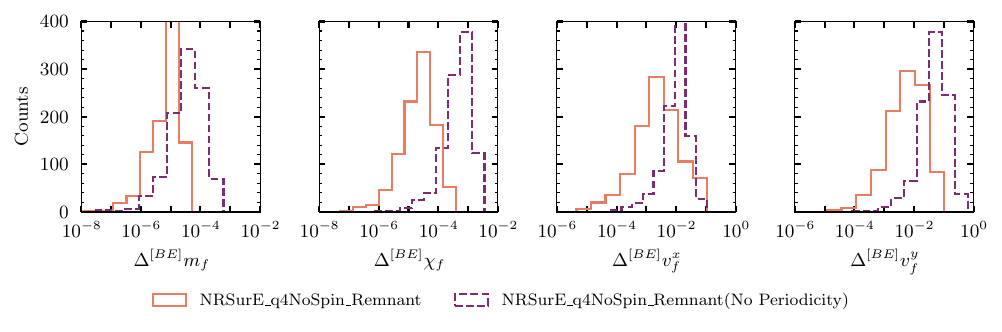}
      \caption{Histogram of the boundary relative errors of \RemModel and \texttt{NRSurE\_q4NoSpin\_Remnant(No Periodicity)}, which respectively include and omit periodicity padding. For each model, we evaluate the surrogate at randomized values of $q$ and $e_{-1000M}$ and compute the boundary error, $\Delta^{[BE]}$, defined in Eq.~\eqref{eq:boundary_err} as the relative difference between evaluations at $\ell_{-1000M}=0$ and $2\pi$. The smaller boundary errors for \RemModel show that periodicity padding described in Sec.~\ref{subsubsection:periodicity_degeneracy} improves the model's ability to capture the periodic dependence on $\ell_{-1000M}$.}
      \label{fig:RemnantSurrogateBoundaryErrHist}
\end{figure*}

We next assess how well the training procedure enforces periodicity in $\ell$. Figure~\ref{fig:RemnantSurrogatePeriodicityCheck_Radial} illustrates the effect of imposing periodicity during training. Each radial plot shows model evaluations at fixed $q=1.5$ and $e_{-1000M}=0.15$, with $\ell_{-1000M}$ as the polar coordinate and a normalized remnant quantity as the radial coordinate. The normalization is defined by

\begin{align}
\delta X = \frac{X - \underset{\ell}{\min}(X_{q=2.0, e=0.15})}{\underset{\ell}{\max}(X_{q=2.0, e=0.15}) - \underset{\ell}{\min}(X_{q=2.0, e=0.15})} \,,
\label{eq:rescaled_X_variation}
\end{align}

where $X \in \{m_f, \chi_f, v_f^x, v_f^y\}$. This rescaling makes differences near the boundaries easier to see. We find that the evaluations at $\ell_{-1000M} = 0$ and $2\pi$ agree more closely when periodicity padding is included. This is further supported by Fig.~\ref{fig:RemnantSurrogateBoundaryErrHist}, which shows histograms of the boundary errors, defined as the relative difference between evaluations at $\ell_{-1000M} = 0$ and $2\pi$, for randomized values of $q$ and $e_{-1000M}$. Here, boundary errors are defined as,

\begin{align}
      \Delta^{[BE]} X_{q, e}
      = \left | \frac{X_{q,e,\ell=2\pi} - X_{q, e,\ell=0}}{X_{q, e, \ell=0}} \right | \,,
      \label{eq:boundary_err}
\end{align}

where each $X$ evaluation is separately done using a remnant model built with (\RemModel) and without (\texttt{NRSurE\_q4NoSpin\_Remnant(No Periodicity)}) periodicity enforced. The relative difference of evaluations at $\ell_{-1000M} = \{0, 2\pi\}$ radians (i.e., boundary error) decreases with periodicity padding implemented, indicating that the padded model more faithfully captures the periodic dependence on $\ell_{-1000M}$. In Eqs.~\eqref{eq:boundary_err} and \eqref{eq:rescaled_X_variation} we omit the parameter reference times for brevity.

\subsection{\DynModel}

\begin{figure}[h]
      \includegraphics{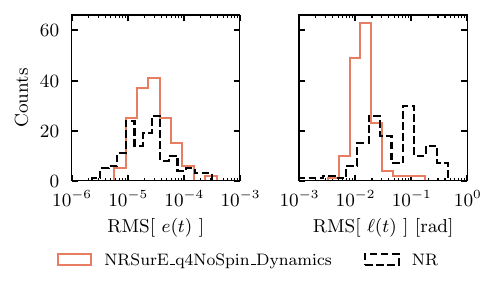}
      \caption{Error histograms of \DynModel predictions. Each histogram represents the distribution of RMS error in (left to right) eccentricity ($e(t)$) and mean anomaly ($\ell(t)$) surrogates from K-fold cross-validation (orange) and NR resolution error (dashed black). The validation errors are less than or comparable to the NR resolution errors, indicating that our model is comparable in accuracy to the NR simulations themselves.}
      \label{fig:AuxSurrogateValidationHistogram}
\end{figure}

\begin{figure}[h]
      \includegraphics{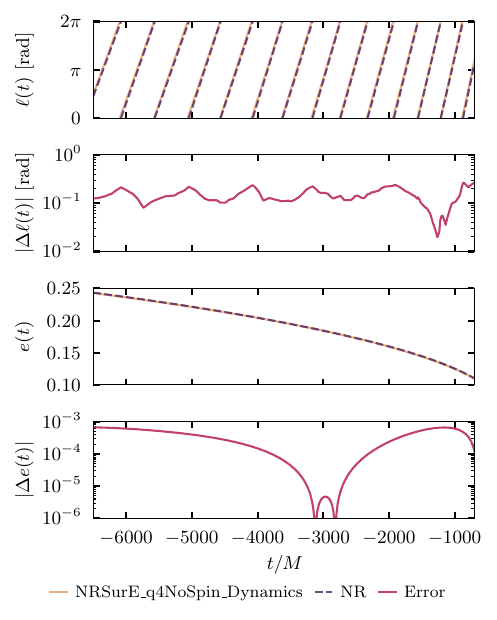}
      \caption{\DynModel compared with NR for the validation-set cases with the largest RMS errors in $\ell(t)$ and $e(t)$. {\bf Top two panels}: The case with the largest RMS error in $\ell(t)$, corresponding to a parameter value of $q=1.8$, $e_{-3000M}=0.001$, and $\ell_{-1200M}=0.067\pi$ radians. The top panel shows $\ell(t)$ from the surrogate (orange solid) and NR (purple dashed), plotted modulo $2\pi$, and the second panel shows the corresponding absolute error $|\Delta \ell(t)|$. {\bf Bottom two panels}: The case with the largest RMS error in $e(t)$, at $q=1.0$, $e_{-3000M}=0.184$, and $\ell_{-1200M}=1.4\pi$ radians. The third panel shows $e(t)$ from the surrogate and NR, and the bottom panel shows the corresponding absolute error $|\Delta e(t)|$. The RMS errors are $1.54\times10^{-1}$ radians for $\ell(t)$ and $3.97\times10^{-4}$ for $e(t)$. The worst case for $\ell(t)$ occurs at very low eccentricity, where the mean anomaly is difficult to determine because the system is nearly quasicircular, whereas the worst case for $e(t)$ occurs near the boundary of the parameter space.
      }
      \label{fig:AuxSurrogateWorstCase}
\end{figure}

Figure~\ref{fig:AuxSurrogateValidationHistogram} shows the out-of-sample errors for \DynModel. Because the purpose of the model is to predict the time evolution of the eccentricity and mean anomaly across the full domain, the root-mean-squared (RMS) error is a natural metric, as it provides a useful summary of the typical pointwise error over time. Each histogram shows the distribution of RMS errors for the eccentricity $e(t)$ and mean anomaly $\ell(t)$ obtained from $20$-fold cross validation. We also show the corresponding NR resolution errors for comparison. The validation errors are smaller than or comparable to the NR resolution errors, indicating that the model is nearly comparable in accuracy to the underlying NR data.

Figure~\ref{fig:AuxSurrogateWorstCase} shows model evaluations at the validation-set parameters with the largest RMS errors for $e(t)$ and $\ell(t)$, considered separately. Even in these worst-case examples, the model captures the overall behavior of the NR data well. For $\ell(t)$, the worst case occurs at low eccentricity. In this regime, although $\ell(t)$ remains well defined, it is more difficult to extract accurately because the system is close to quasi-circular and the peaks (periastron passages) and troughs (apastron passages) in the amplitude or frequency become less pronounced~\cite{Shaikh:2023ypz}. The corresponding RMS error is $1.54\times10^{-1}$ radians. For $e(t)$, the worst case occurs at higher eccentricity and $q=1$, near the boundary of the parameter space. 
The corresponding RMS error is $3.97\times10^{-4}$.

We find that the model accuracy is not especially sensitive to the choice of reference times used for parameterization. We therefore fix these reference times to $t_{\mathrm{ref}}=-3000M$ for eccentricity and $t_{\mathrm{ref}}=-1200M$ for mean anomaly, so that they match the parameterization used by the waveform surrogate model \texttt{NRSurE\_q4NoSpin\_22}~\cite{Nee:2025nmh}.

\section{Applications}
\label{sec:applications}

In the previous sections, we described two surrogate models: \RemModel for remnant properties and \DynModel for the evolution of eccentricity and mean anomaly. These models have several applications. The remnant model can be used to study eccentric hierarchical BBH mergers, test eccentric BBH population models, and quantify quasinormal modes for eccentric mergers, among other uses. Applications of the dynamics model include constructing eccentric inspiral-merger-ringdown waveform surrogates through local parametric fitting and reparameterizing existing eccentric models at different reference times.

In this section, we present two simple applications of these surrogate models. In Sec.~\ref{subsubsection:eval_remnant_properties}, we evaluate the trained remnant model to examine whether it smoothly captures the dependence of remnant properties on the input parameters, particularly the effects of eccentricity. In Sec.~\ref{sec:remnant_oscillations}, we use \RemModel to study the oscillations in remnant properties as a function of eccentricity~\cite{Wang:2023vka, Wang:2024jro, Wang:2023wol, Radia:2021hjs, Carullo:2023kvj}. In Sec.~\ref{sec:remnant_from_dynamics}, we show how \DynModel can be used to parameterize \RemModel for applications in which remnant properties must be predicted from initial parameters defined at different reference times.

\subsubsection{Evaluating \RemModel for remnant properties}
\label{subsubsection:eval_remnant_properties}

\begin{figure}[h!]
      \includegraphics{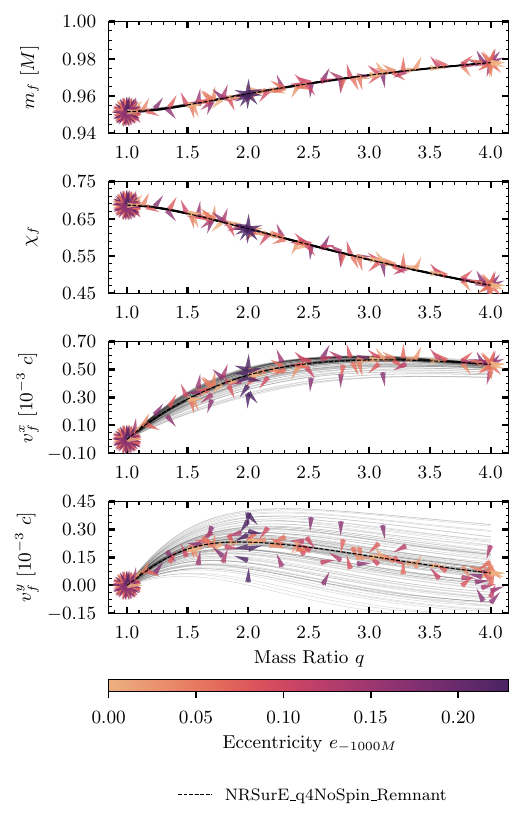}
      \caption{\RemModel evaluations at varying $q$, with $e_{-1000M}=0.$ and $\ell_{-1000M}=\pi$ radians (randomized $e_{-1000M}$ and $\ell_{-1000M}$) as the black dashed line (solid gray lines). Each arrow marker represents the value of remnant properties (top to bottom: remnant mass, remnant spin, remnant kick velocity $x-y$ components) for each point in the training set, like in Fig.~\ref{fig:RemnantProperties}.  It is evident that the model successfully captures the leading order dependence in $q$. For the remnant kick velocities, the effect of $e$ and $\ell$ can be seen as small oscillations around the leading order dependence in the plot, and the model captures those effects as well.}
      \label{fig:RemnantSurrogateLeadingOrder}
\end{figure}

Figure~\ref{fig:RemnantSurrogateLeadingOrder} shows model evaluations as a function of $q$ overlaid on the data from Fig.~\ref{fig:RemnantProperties}. The black dashed curve denotes the model evaluated at $e_{-1000M}=0$ and $\ell_{-1000M}=\pi$. The model clearly captures the leading-order dependence on $q$. For the remnant kick velocities, the data show a larger spread about this leading-order trend. The solid gray curves, which correspond to evaluations on a uniform random grid with $e_{-1000M}\leq 0.23$ and $\ell_{-1000M}\in [0,2\pi)$, show the additional eccentricity- and anomaly-dependent variations.

The model can also be used to study how $e$ and $\ell$ enhance or suppress the kick relative to the corresponding quasi-circular system. For example, previous work~\cite{Sperhake:2019wwo} found that this variation can reach $25$\% in the superkick configuration, which occurs in spinning BBH systems. In the non-spinning parameter space we consider here, our model predicts the eccentricity(and mean anomaly)-induced variation ranges from about $2$\% to $10$\%, depending on the mass ratio.

\subsubsection{Oscillatory behavior in remnant properties}
\label{sec:remnant_oscillations}
\begin{figure*}[tb!]
      \includegraphics{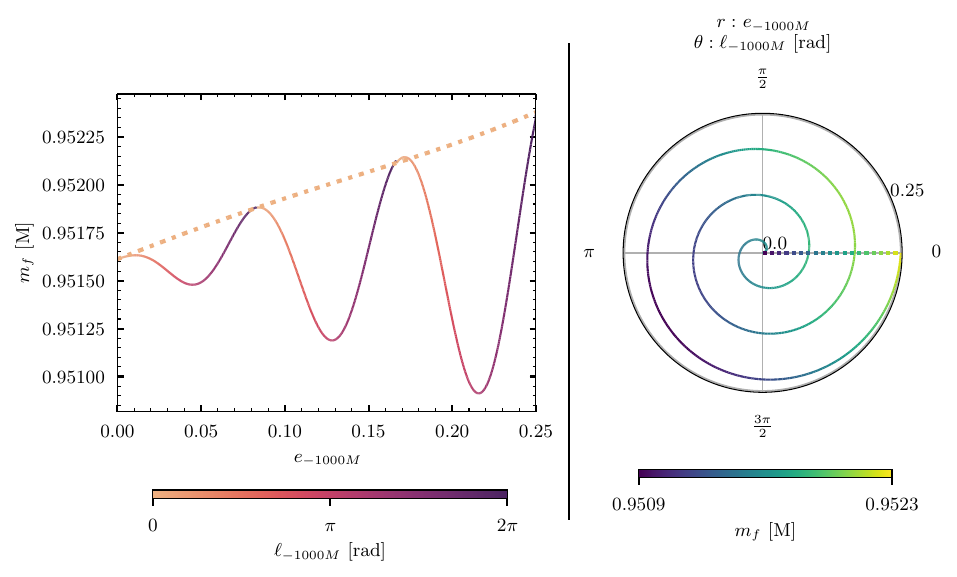}
      \caption{\RemModel evaluation at $q=1$ along the two one-dimensional paths through the $e_{-1000M}\otimes\ell_{-1000M}$ space defined in Eq.~\eqref{eq:slices}. The left panel shows $m_f$ as a function of $e_{-1000M}$, with color indicating $\ell_{-1000M}$ along each path. The right panel shows the same paths in polar coordinates, with $e_{-1000M}$ as the radial coordinate, $\ell_{-1000M}$ as the polar angle, and $m_f$ shown by the color scale. The dotted curve corresponds to a path where the eccentricity increases while the mean anomaly is held fixed (path $c_2$ in Eq.~\eqref{eq:slices}) while the solid curve corresponds to a path (path $c_1$ in Eq.~\eqref{eq:slices}) where the eccentricity increases while the mean anomaly varies simultaneously and winds three times through $[0,2\pi)$. When $\ell_{-1000M}$ is held fixed, $m_f$ varies smoothly with $e_{-1000M}$ and does not oscillate. The oscillations arise only along the spiral path, where $\ell_{-1000M}$ varies together with $e_{-1000M}$. This shows that the oscillatory structure is driven by the anomaly, while the eccentricity sets its overall scale.}
      \label{fig:RemnantSurrogateRITOsc}
\end{figure*}

Eccentricity-induced oscillations have been reported in remnant properties when they are plotted as functions of eccentricity alone~\cite{Wang:2023vka, Wang:2024jro, Wang:2023wol, Radia:2021hjs, Carullo:2023kvj}. Subsequent work~\cite{Nee:2025zdy} showed that these oscillations arise from neglecting the dependence on mean anomaly, thereby projecting an underlying two-dimensional dependence on eccentricity and mean anomaly onto a single eccentricity variable defined at fixed separation. We now use \RemModel to show that it captures this two-dimensional structure and to illustrate how the oscillations arise from the mean-anomaly dependence.

To do so, we consider two one-dimensional paths through the $e_{-1000M} \otimes \ell_{-1000M}$ space at fixed $q=1$. Along the first path, the eccentricity increases while the mean anomaly varies simultaneously and winds three times through $[0,2\pi)$. Along the second path, the eccentricity again increases, but the mean anomaly is held fixed. These paths are
\begin{align}\label{eq:slices} 
c_1(\lambda):\ & e_{-1000M} = 0.25\lambda, \ell_{-1000M} = 6\pi\lambda \mod 2\pi \text{ radians } \nonumber \\ 
c_2(\lambda):\ & e_{-1000M} = 0.25\lambda, \ell_{-1000M}=0 \text{ radians } ,
\end{align}
where $\lambda \in [0, 1]$. Figure~\ref{fig:RemnantSurrogateRITOsc} shows \RemModel evaluated along these two paths at $q=1$, with the solid and dotted curves corresponding to $c_1$ and $c_2$, respectively. Along $c_2$, where the mean anomaly is fixed, the remnant mass $m_f$ varies smoothly with eccentricity and does not exhibit oscillations. Along $c_1$, by contrast, oscillations appear when $m_f$ is plotted against $e_{-1000M}$. These oscillations are therefore not caused by eccentricity alone, rather they arise because $\ell_{-1000M}$ varies simultaneously with $e_{-1000M}$. Put differently, the oscillatory structure is driven by the mean anomaly, while the eccentricity sets its overall scale. We find similar behavior for the other remnant quantities as well. This agrees with Ref.~\cite{Nee:2025zdy}, which reached the same conclusion directly from NR simulations, and further supports the interpretation that the observed oscillations result from following a particular one-dimensional path through an underlying two-dimensional parameter space.

\subsubsection{Using \DynModel to parameterize \RemModel}
\label{sec:remnant_from_dynamics}

Here we display the effectiveness of using the dynamics model to accurately map two different choices of reference times for $e$ and $\ell$ to parameterize the remnant models. Consequently, this also shows that the remnant model can be parameterized at different reference times than the ones used to build the model, by using the dynamics model to map to different times. The flow of information is as follows:
\begin{enumerate}
      \item Given a parameter value $(e_{-3000M},\ \ell_{-1200M},\ q)$ in dataset, evaluate \DynModel for $e(t)$ and $\ell(t)$.
      \item Extract $e^{*}_{-1000M}$ and $\ell^{*}_{-1000M}$ from the $e(t)$ and $\ell(t)$.
      \item Evaluate \RemModel at $e^{*}_{-1000M}$ and $\ell^{*}_{-1000M}$ and compare with true NR values.
\end{enumerate}

\begin{figure*}[tb!]
      \includegraphics{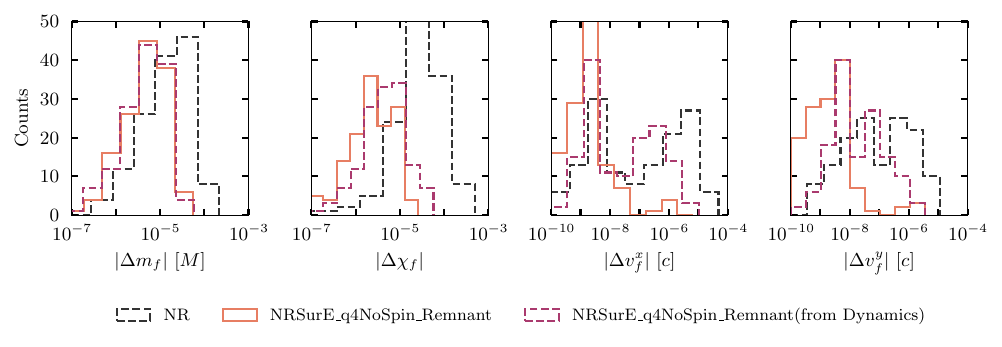}
      \caption{Histograms of \RemModel model errors when parameterized using \DynModel to map from $(e_{-3000M}, \ell_{-1200M})$ to $(e_{-1000M}, \ell_{-1000M})$ (purple). These errors are compared to the $K$-fold cross-validation errors of \RemModel (orange) and NR resolution errors (dashed black). Each histogram represents the distribution of absolute errors in (left to right) remnant mass, remnant spin, remnant kick velocity components. All three errors are comparable, indicating that the dynamics model can be effectively used to parameterize the remnant model at different reference times.}
      \label{fig:RemnantFromAuxSurrogateValidationHistogram}
\end{figure*}

Figure~\ref{fig:RemnantFromAuxSurrogateValidationHistogram} shows that the errors from this procedure are comparable to the NR resolution and model errors, indicating that the dynamics model can be effectively used to parameterize the remnant model at different reference times.

\section{Conclusions}
\label{sec:conclusions}

We have presented two accurate surrogate models for eccentric, nonspinning, unequal-mass black hole binaries: \RemModel, which predicts the remnant mass, spin magnitude, and recoil velocity, and \DynModel, which models the inspiral evolution of the eccentricity and mean anomaly. The models are trained on 203 and 151 NR simulations, respectively, spanning mass ratios $q \leq 4$ and eccentricities $e \leq 0.23$, with eccentricity defined at $t=-1000M$ relative to the waveform's peak amplitude (cf. Eq.~(24) of Ref.~\cite{Blackman:2017dfb}). The NR simulations were performed with \texttt{SpEC}~\cite{Kidder_2000}. The remnant quantities were extracted using CCE through the CCE module in \texttt{SpECTRE}~\cite{spectrecode,Moxon:2020gha,Moxon:2021gbv}, while the dynamical quantities were obtained using the prescription of Ref.~\cite{Shaikh:2023ypz}. To improve the treatment of the periodic mean-anomaly dependence, \RemModel was trained with padding in the $\ell$ direction as described in Sec.~\ref{subsubsection:periodicity_degeneracy}, while for \DynModel, we found that analogous augmentation was unnecessary.

\RemModel was constructed using GPR with the methods and settings of Ref.~\cite{Varma:2018aht}, implemented through scikit-learn~\cite{scikit-learn}, while \DynModel was built using the time-domain surrogate procedure of Ref.~\cite{Field:2013cfa}. We assessed both models through $K$-fold cross-validation and compared the resulting validation-error distributions with NR resolution errors across the dataset. In both cases, the surrogate errors are comparable to the NR resolution errors, indicating that the models achieve accuracy close to that of the underlying simulations over their training domains. For the remnant quantities computed with \RemModel, the median errors are $\approx 9 \times 10^{-6} M$ for $m_f$, $\approx 8 \times 10^{-6}$ for $\chi_f$, $\approx 4 \times 10^{-7} c$ for $v_f^x$, $\approx 4 \times 10^{-7} c$ for $v_f^y$, $\approx 7 \times 10^{-7}c $ for the kick speed, and $\approx 3 \times 10^{-3}$ radians for the kick angle.

For \RemModel, we also quantified the systematic error that arises when a quasicircular remnant model is used to predict remnants of eccentric systems. Specifically, we applied \texttt{NRSur3dq8Remnant}~\cite{Varma:2018aht} to eccentric binaries and compared the resulting errors with both the surrogate validation errors and the NR resolution errors. The quasicircular-model errors are between $1$ and $2$ orders of magnitude larger than both (see Fig.~\ref{fig:RemnantSurrogateValidationHistogram}), demonstrating that eccentricity must be modeled explicitly for accurate remnant predictions. 

These models have several applications in GW physics. \RemModel can be used to study eccentric hierarchical BBH mergers, test eccentric BBH population models, and quantify quasinormal modes for eccentric mergers. Applications of \DynModel include constructing eccentric inspiral-merger-ringdown waveform surrogates through local parametric fitting and reparameterizing existing eccentric models at different reference times. In this paper, we explored two applications. First, we used \RemModel to clarify the origin of the oscillations in remnant properties that have been reported in recent years when those quantities are plotted as functions of eccentricity alone~\cite{Wang:2023vka, Wang:2024jro, Wang:2023wol, Radia:2021hjs, Carullo:2023kvj}. By evaluating the model along different one-dimensional paths through the underlying two-dimensional $(e,\ell)$ parameter space, we showed that these oscillations do not arise from the eccentricity by itself. Rather, they appear when the mean anomaly varies simultaneously with the eccentricity, with the eccentricity primarily setting the overall scale of the variation. Second, we showed how \DynModel can be used to parameterize \RemModel, which is useful when one wishes to predict remnant properties from initial $(e,\ell)$ parameters defined at other reference times.

Our models \RemModel and \DynModel will be implemented in the public Python modules \texttt{surfinBH}~\cite{vijay_varma_2018_1435832} and \texttt{GWSurrogate}~\cite{Field:2025isp}, respectively. In the future, we plan to extend these models to higher mass ratios, spinning systems, and larger eccentricities as additional NR simulations of eccentric black hole binaries become available.

\begin{acknowledgments}
      \texttt{SpECTRE} uses \texttt{Charm++}/\texttt{Converse}~\cite{laxmikant_kale_2020_3972617}, which was developed by the Parallel Programming Laboratory in the Department of Computer Science at the University of Illinois at Urbana-Champaign.\cite{laxmikant_kale_2020_3972617}
      K.M. is supported by NASA through the NASA Hubble
      Fellowship grant \#HST-HF2-51562.001-A awarded by
      the Space Telescope Science Institute, which is operated
      by the Association of Universities for Research in Astronomy, Incorporated, under NASA contract NAS5-26555, the National Science Foundation under Grants No. PHY-2407742, No. PHY-2207342,
      and No. OAC-2209655, and by the Sherman Fairchild Foundation.
      V.V.~acknowledges support from NSF Grant No. PHY-2309301.
      S.F.\ acknowledges support from NSF Grants No. AST-2407454 and PHY-2110496.
      The work of LCS was supported by NSF CAREER Award PHY–2047382 and a Sloan Foundation Research Fellowship.
      This material is based upon work supported by the National Science Foundation under Grants No. PHY-2407742, No. PHY-2308615, and No. OAC-2513338, and by the Sherman Fairchild Foundation at Cornell.
      A. R.-B. is supported by the Veni research programme which is (partly) financed by the Dutch Research Council (NWO) under the grant VI.Veni.222.396; acknowledges support from the the Universitat de les Illes Balears (UIB) and the Spanish Agencia Estatal de Investigación grant PID2022-138626NB-I00 funded by MICIU/AEI/10.13039/501100011033 and the ERDF/EU,  PID2024-157460NA-I00; and the Spanish Ministerio de Ciencia, Innovación y Universidades (Beatriz Galindo, BG23/00056), co-financed by UIB.
      P.K. acknowledges support of the Department of Atomic Energy, Government of India, under project no. RTI4019; and by the Ashok and Gita Vaish Early Career Faculty Fellowship at the International Centre for Theoretical Sciences.
      H.\,R.\,R.\ acknowledges financial support provided under the European Union's H2020 ERC Advanced Grant ``Black holes: gravitational engines of discovery'' grant agreement no.~Gravitas–101052587. Views and opinions expressed are however those of the authors only and do not necessarily reflect those of the European Union or the European Research Council.  Neither the European Union nor the granting authority can be held responsible for them.
      This work was partly supported by UMass Dartmouth's Marine and Undersea Technology (MUST) research program funded by the Office of Naval Research (ONR) under grant No.\ N00014-23-1-2141.
      This work is supported by the Sherman Fairchild Foundation, the National Science Foundation under Grants No.~PHY-2309211, No.~PHY-2309231, and No.~OAC-2513339 at Caltech, and NASA award No. 80NSSC26K0340.
      A.C. acknowledges support from PHY-2208014, AST-2219109, Nicholas and Lee Begovich, and the Dan Black Family Trust.
      This material is based upon work supported by the National Science Foundation under Grants No.~PHY-2309211; No.~PHY-2309231; and No.~OAC-2513339 at Caltech, and No.~PHY-2407742; No.~PHY-2207342; and No.~OAC-2513338 at Cornell. Any opinions, findings, and conclusions or recommendations expressed in this material are those of the author(s) and do not necessarily reflect the views of the National Science Foundation. This work was supported by the Sherman Fairchild Foundation at Caltech and Cornell.
      This material is based upon work supported by NSF's LIGO Laboratory which is a major facility fully funded by the NSF.

\end{acknowledgments}

\section*{References}
\bibliography{References.bib}

\appendix

\section{Duplication across $\ell$ boundaries to enforce periodicity in mean anomaly}
\label{section:proof_of_principle_duplication}

The mean anomaly $\ell$ is a cyclic parameter, so $\ell=0$ radians and $\ell=2\pi$ radians correspond to the same physical configuration. In Sec.~\ref{subsubsection:periodicity_degeneracy}, we described a technique to help guide the model toward this expected periodic behavior. The main idea is to enlarge the mean-anomaly domain beyond its physical interval $[0,2\pi)$ and duplicate points into the extended region, assigning them the same remnant properties as the original data.

\begin{figure}[H]
      \centering
      \includegraphics[width=\columnwidth]{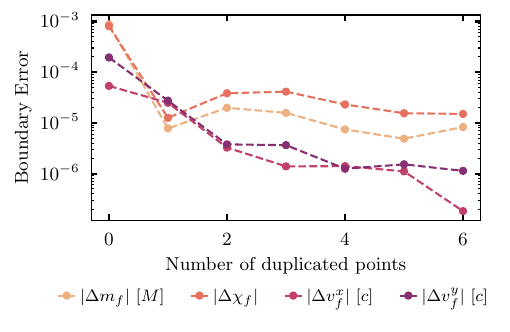}
      \caption{Boundary error (see Eq.~\ref{eq:boundary_err_appendix}) for toy model surrogates as a function of the number of duplicated points added to the training set. Each point added was increasingly farther from the $\ell_{-1000M}$ boundaries. It is evident that when duplicated points are added, the differences decrease, indicating that the model is learning the periodic nature of the mean anomaly.}
      \label{fig:RemnantSurrogateDeltaEvalVsMeanAno-combined}
\end{figure}

As a proof of principle, we demonstrate the effectiveness of duplication across the mean-anomaly boundaries by training a GPR model on a one-dimensional dataset of six NR simulations with $q=2$, $e_{-1000M}=0.2283 \pm 0.0005$, and $\ell_{-1000M}$ in $[0,2\pi)$ radians. This dataset ensures that the remnant properties vary only with mean anomaly, allowing us to test the efficacy of the duplication strategy more cleanly.

We build a sequence of surrogate models by duplicating increasingly more points, resulting in a total of six surrogates, and compute a boundary error defined as
\begin{equation}
      \text{Boundary Error } = \left | X_{\ell=2\pi} - X_{\ell=0} \right | \, ,
      \label{eq:boundary_err_appendix}
\end{equation}

where $X \in \{m_f, \chi_f, v_f^x, v_f^y\}$. The boundary error is our main diagnostic for assessing the model's ability to capture periodicity, since it should vanish if the model is exactly periodic at $\ell=0$ and $\ell=2\pi$.

\begin{figure}[H]
      \centering
      \includegraphics[width=\columnwidth]{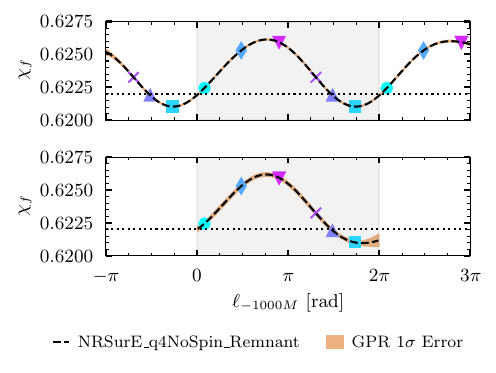}
      \caption{The top (bottom) panel shows the toy-model surrogate for the remnant spin $\chi_f$ evaluated across the mean-anomaly domain with (without) periodicity encouraged through duplication. The markers denote the NR data points, together with their duplicates when present, the black curve denotes the model trained on the original dataset, and the orange shaded region denotes the GPR $1\sigma$ error estimate. The horizontal dotted line marks the model prediction at $\ell_{-1000M}=0$ radians, which by periodicity should equal the value at $\ell_{-1000M}=2\pi$ radians. The top panel shows that duplication enables the surrogate to capture the expected periodic behavior and improves the predictions near the domain boundaries, while the bottom panel shows that without duplication the model fails to do so. Our data duplication procedure also noticeably reduces the GPR uncertainty near the $\ell=2\pi$ boundary.}
      \label{fig:RemnantSurrogateEvalVsMeanAno-combined}
\end{figure}

Figure~\ref{fig:RemnantSurrogateDeltaEvalVsMeanAno-combined} shows that the boundary error decreases as more points are duplicated. It also shows that the model trained on the original dataset, with no duplication, fails to capture the periodic nature of the mean anomaly, leading to a discontinuity and larger GPR $1\sigma$ error near the boundaries. In contrast, the model trained on a maximally duplicated dataset accurately captures the periodicity, provides consistent predictions across the full extended domain, and further reduces the GPR $1\sigma$ error near the boundaries. Figure~\ref{fig:RemnantSurrogateEvalVsMeanAno-combined} provides a direct comparison of the model's prediction for the remnant spin with and without duplication.

For the remnant model developed in this paper, \RemModel, we duplicate data points only after extending the mean-anomaly domain by a size $\delta$, which is smaller than the full $2\pi$ extension shown in Fig.~\ref{fig:RemnantSurrogateEvalVsMeanAno-combined}. This is because the build-time and evaluation-time costs of GPR scale as $\mathcal{O}(N^3)$ and $\mathcal{O}(N^2)$, respectively, where $N$ is the number of data points. Since our full dataset contains $203$ NR simulations, duplicating too many points would make both training and evaluation more expensive. Finally, all duplication is carried out at surrogate build time so that no duplicated point enters the validation set, ensuring that the $K$-fold cross-validation errors are not artificially lowered.

\end{document}